\documentclass[conference]{IEEEtran}
\IEEEoverridecommandlockouts
\usepackage{amsmath,amssymb,amsfonts}
\usepackage{algorithmic}
\usepackage{graphicx}
\usepackage{textcomp}
\usepackage{xcolor}
\usepackage{cite}
\usepackage{float}
\usepackage{stfloats}
\usepackage{flushend}
\def\BibTeX{{\rm B\kern-.05em{\sc i\kern-.025em b}\kern-.08em
    T\kern-.1667em\lower.7ex\hbox{E}\kern-.125emX}}
\begin{document}

\title{Psychlysis: Towards the Creation of a Questionnaire-based Machine Learning Tool to Analyze States of Mind}

\author{\IEEEauthorblockN{Hemakshi Jani}
\IEEEauthorblockA{\textit{School of Computer Science} \\
\textit{University of Windsor}\\
Windsor, Ontario, Canada \\
jani32@uwindsor.ca}\\
\IEEEauthorblockN{Rahul Bhadja}
\IEEEauthorblockA{\textit{School of Computer Science} \\
\textit{University of Windsor}\\
Windsor, Ontario, Canada \\
bhadjar@uwindsor.ca}
\and
\IEEEauthorblockN{Mitish Karia}
\IEEEauthorblockA{\textit{School of Computer Science} \\
\textit{University of Windsor}\\
Windsor, Ontario, Canada \\
karia2@uwindsor.ca}\\
\IEEEauthorblockN{Aznam Yacoub}
\IEEEauthorblockA{\textit{School of Computer Science} \\
\textit{University of Windsor}\\
Windsor, Ontario, Canada \\
aznam.yacoub@uwindsor.ca}
\and
\IEEEauthorblockN{Meet Gohil}
\IEEEauthorblockA{\textit{School of Computer Science} \\
\textit{University of Windsor}\\
Windsor, Ontario, Canada \\
gohil41@uwindsor.ca}\\
\IEEEauthorblockN{Shafaq Khan}
\IEEEauthorblockA{\textit{School of Computer Science} \\
\textit{University of Windsor}\\
Windsor, Ontario, Canada \\
shafaq.khan@uwindsor.ca}
}

\maketitle

\begin{abstract}
This paper describes the development of Psychlysis, a \textbf{work-in-progress} questionnaire-based machine learning application analyzing the user's current state of mind and suggesting ways to improve their mood using Machine Learning. The application utilizes the OCEAN model to understand the user's personality traits and make customized suggestions to enhance their well-being. The proposed application focus on improving the user's mood rather than just detecting their emotions. Preliminary results of the model are presented, showing the potential of the application in predicting the user's mood and providing personalized recommendations. The paper concludes by highlighting the potential benefits of such an application for various societal segments, including doctors, individuals, and mental health organizations, in improving emotional well-being and reducing the negative impact of mental health issues on daily life. 
\end{abstract}


\section{Introduction and Motivation}
\label{sec-introduction}
The American Psychological Association defines emotion as 
\begin{quote}
    a complex reaction pattern, involving experiential, behavioural, and physiological elements by which individuals attempt to deal with a personally significant matter or event \cite{APADictionary:Emotion:2015}.
\end{quote} 
Human cognitive functions such as perception, attention, learning, memory, reasoning, and problem-solving are all significantly influenced by emotions. Emotion has a particularly potent impact on attention, modifying its selectivity in particular and influencing behaviour and action motivation. 

Mental and social conditions, such as loneliness, have a significant impact on emotional dysregulation and life. Social isolation has exacerbated these issues, especially during the COVID-19 pandemic, causing increased levels of mental illnesses \cite{Choi:10.3390/ijerph17103740:2022, Roy:10.1016/j.ajp.2020.102083:2020, Velotti:10.3389/fpsyt.2020.581494:2021, Zhang:10.3389/fpsyg.2021.681091:2021}. According to World Health Organization \cite{WHO:MentalDisorders:2022}, 970 million people worldwide were living with mental disorders in 2019, with anxiety and depressive disorders being the most common. Positive thinking, defined as an overall attitude reflected in thinking, behaviour, feeling, speaking, thoughts, words, and images leading to growth, expansion, and success \cite{Naseem:2010}, is good for the immune system, anxiety reduction, and increasing positive emotions such as happiness \cite{Eagleson:10.1016/j.brat.2015.12.017:2016, Macleod:10.1002/(SICI)1099-0879(200002)7:1<1::AID-CPP228>3.0.CO;2-S:2000, Shokrpour:10.4103/jehp.jehp_1120_20:2021}. Positive thinking aims to use mood-altering techniques like meditation to help users break free from ingrained patterns and habits. 

In this paper, we discuss the idea of the development of a software-based system helping in mood and state of mind identification to assist in removing unfavorable mental conditions that negatively impact daily activities. Psychlysis is a \textbf{work-in-progress} web-based application using a Machine Learning (ML) algorithm designed to analyze the user's current state of mind and provide them with personalized recommendations to improve their mood. The application uses the five-factor OCEAN model \cite{McCrae:10.1111/j.1467-6494.1992.tb00970.x:1992} to understand the user's personality traits and tailor recommendations to their specific needs. It helps identify negative thoughts and provides the user with a list of techniques to break their negative feelings. The software seeks also to store relevant past mood history to help doctors in mental health evaluation. Ultimately, Psychlysis aims to provide personalized recommendations to enhance the user's daily life, reduce the negative impact of mental health issues, and aid in emotional well-being. Section \ref{sec-related-work} briefly summarizes existing ML-based approaches for state-of-mind analysis. Key points of the architecture of Psychlysis are provided in Section \ref{sec-proposed-model}. Preliminary results and current limitations are finally discussed in Section \ref{sec-results-limits}.

\section{Related Work}
\label{sec-related-work}
The growth of Machine Learning (ML) has allowed interesting progress and advances in Computational Psychology since the last decade \cite{Brown:10.5555/3016387.3016562:2016}. Researchers have started trying to use ML techniques also in state-of-mind analysis and mood enhancing\cite{Jamisola:2016, Jamisola:2021}. These techniques can be classified according to the modalities employed for mood detection. Facial expression analysis is a well-used approach for automatic detection of emotion. \cite{Dubey:2016}. Manjula et al. \cite{Manjula:10.1007/978-981-15-7990-5_33:2021} use this approach to introduce a mobile application to identify facial expressions, associate a corresponding mood, and uplift the user's mindset by providing adequate responses. The authors employ various open-source machine learning libraries like Weka, OpenCV, Dlib, and TensorFlow Lite to classify facial expressions with high accuracy. However, due to memory limitations, the accuracy of the application may be compromised: for instance, the user needs to ensure proper lighting and positioning for optimal results. Apart from these technical considerations, we can also oppose that facial recognition allows only emotional recognition and not mood recognition. Indeed, mood and emotions are distinct phenomena \cite{Beedie:2005}: mood is a complex long-lasting mental state which is not associated with a specific facial expression, contrary to emotions, which are already not always reflected in physical appearance.
Correlate Facial Expression Recognition (FER) with Situational Awareness (SA) \cite{Aguinaga:10.3390/app10051736:2020} is a way to overcome this limit. SA \cite{Stanton:10.1080/00140139.2017.1278796:2017} is defined as the ability to understand an environment and its evolution with respect to time and multiple factors. Aguiñaga et al. \cite{Aguinaga:10.3390/app10051736:2020} propose a strategy to analyze the cognitive effects caused by work activities using video analysis and measure the levels of comfort and mental stability of a worker in their working environment. Pictures are captured continuously using cameras placed in the workplace. Emotions are analyzed using a Convolutional Neural Network (CNN) coupled with the Extreme Sparse Learning (ESL) algorithm \cite{Shojaeilangari:10.1109/TIP.2015.2416634:2015}, while situation awareness is modeled using Endsley’s model \cite{Endsley:/10.1002/0470048204.ch20:2006}. This study ends by demonstrating the main issues in mood analysis: fully modeling and understanding all the probable factors require a multidisciplinary effort, which is generally hard to implement. Moreover, it is difficult to identify parameters outside of the experimental conditions. For instance, in their study, the authors choose carefully the position of the cameras in order not to interfere with the employees' behaviour. This poses the problem of the identification of external factors which cause mood modifications in a broader context, and sensors to capture them efficiently, without inducing biases by the experiment itself. 

Electroencephalogram (EEG)-based approaches \cite{Torres:10.3390/s20185083:2020} recognize the difficulty to make the distinction between mood and emotion and propose a multivariate analysis based on EEG signals to improve the accuracy of the emotional state of mind recognition. In \cite{Kumar:10.3390/computers11100152:2022}, the classification of emotions into positive and negative categories is implemented using Bi-Directional Long Short-Term Memory (Bi-LSTM) and considering different brain regions. If accuracy has been improved according to the paper, major drawbacks concern the need to take into account electrode-specific work, which increases the complexity of the computations, and to identify and integrate more features in order to understand how each electrical signal induced by stimuli is associated with a specific emotion. Technically, it implies also the use of wearable sensors, which can hardly be scaled or convenient for daily use. Similarly, Dadebayev et al. conducted a comprehensive analysis of the research done over the last five years regarding the ability of affordable commercial EEG devices in recognizing emotions. Their conclusion is unequivocal: these devices are yet to prove their efficiency in emotion recognition while 'there is no sufficient evidence to declare the optimum~\lbrack EEG-based \rbrack~machine learning algorithm for emotion recognition tasks' \cite[p.~4398]{Dadebayev:10.1016/j.jksuci.2021.03.009:2022}.

\begin{figure}[!b]
    \centering
    \includegraphics[width=2.5cm]{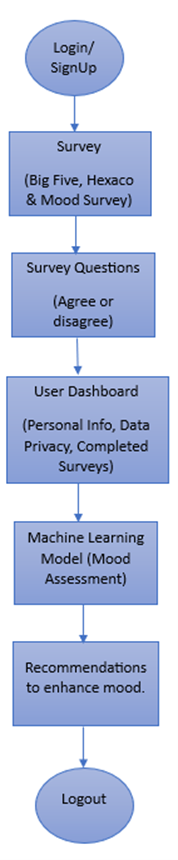}
    \caption{Psychlysis Workflow.}
    \label{fig-psychlysis-arch}
\end{figure}

Creating an \textit{artificial psychologist} to replicate \textit {a human psychologist's understanding} but with potential accuracy concerns related to dishonest responses has been recently investigated. Jamisola \cite{Jamisola:2021} has the idea to create an ML-based algorithm that mimics the expertise of a psychologist by asking thousands of questions. Using a Support Vector Machine (SVM) coupled with an Artificial Neural Network (ANN), the proposed system is able to determine a true or false or degree state of a respondent and identify critical questions which can lead to the precise assessment of one cognitive dimension. In the end, the system is trained to output a judgment given the input collected from the users. As a result, it is able to determine the mental condition of a respondent filling out a questionnaire generated by the system itself.
     
While this approach is likely to replicate biases inherent in questionnaire-based diagnostic examinations, it highlights the potential for machine learning to enhance our understanding of human emotions, in a non-invasive way, and without the need for complex devices. We propose to extend this latter approach and combine it with the OCEAN model in order to identify potential situations which could induce positive and negative thoughts and change user's state of mind \cite{Egges:10.1007/978-3-540-45224-9_63:2003, Giluk:10.1016/j.paid.2009.06.026:2009, OBrien:10.1111/j.1467-6494.1996.tb00944.x:1996}. Therefore, it could enable the possibility to propose actions in order to improve user mood.

\section{Proposed Model}
\label{sec-proposed-model}

\begin{figure*}
    \centering
    \includegraphics[width=13cm]{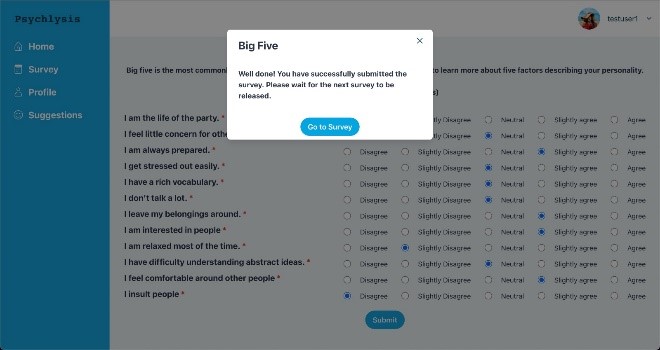}
    \caption{OCEAN Personality Inventory.}
    \label{fig-ocean}
\end{figure*}

\begin{figure*}[t]
    \centering
    \includegraphics[width=13cm]{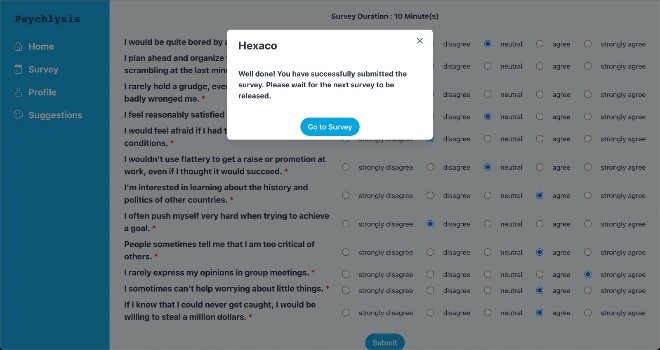}
    \caption{HEXACO Personality Inventory.}
    \label{fig-hexaco}
\end{figure*}

The proposed software (Fig. \ref{fig-psychlysis-arch}) is a web application\footnote{The current version of the source code is available at https://github.com/mitish13/Psychlysis-Model.} which uses several survey forms and an ANN in order to make recommendations to improve user's mood. The application is developed using modern web development frameworks such as React or Angular to ensure data privacy and security. The idea to create a web application comes from two facts: the system can be accessed from any platform, and the underlying learning model can continuously be fed from anonymous data to improve the accuracy of its knowledge about mood detection. In a way, the system is mimicking a human psychologist who would better understand the impacts of recommendations from several patients. The application starts by requesting the user's login or signup information to segregate and protect user responses. Once logged in for the first time, the user is invited to fill out several surveys to capture their personality and lifestyle. These questionnaires replicate OCEAN (Fig. \ref{fig-ocean}) and HEXACO \cite{Ashton:10.1037/0022-3514.86.2.356:2004} (Fig \ref{fig-hexaco}.) inventories. Retrieved answers are then analyzed by an underlying ANN trained on sets of psychological and wellness data correlating personality traits and possible positive/negative thoughts created from literature findings.

\begin{figure}[hb]
    \centering
    \includegraphics[width=7cm]{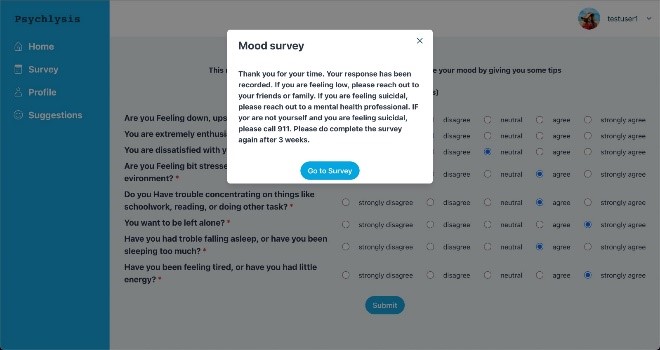}
    \caption{Generated Mood Questionnaire.}
    \label{fig-mood}
\end{figure}

The ANN generates then questions to assess and analyze the mood of the user, and, to enhance it, recommends actions (Fig. \ref{fig-mood}) corresponding to the practices of positive thinking described in the literature. 

\section{Results and Limitations}
\label{sec-results-limits}
Our model is still under development but preliminary results were validated by assessing the adequacy between the actions proposed by the system and the feeling of the user (i.e. if the user thought that the identified mood and proposed actions were relevant, helpful, and accurate). This preliminary validation phase involved 20 students (all of them were foreign students in a university at the graduate level, generally suffering from isolation).

If the model demonstrated a satisfying accuracy (70\% on average), it still suffers from several limitations which need to be addressed in future work:
\begin{enumerate}
    \item Limited Data Availability: The application requires a significant amount of data for it to work efficiently. If there is a lack of data available on the user's personality traits and situational awareness, the model may not provide accurate recommendations. Moreover, the introduction of a feedback loop may create biases if the data used for learning lacks diversity;
    \item User Misinterpretation of Questions: Psychlysis relies on user feedback through questions and answers to understand their personality traits, moods, and mental conditions. If the user misinterprets any questions or provides inaccurate answers, the application may not provide accurate recommendations;
    \item Limited Mood Predictive Power: Although Psychlysis uses mood-altering techniques like meditation to help users break free from ingrained patterns and habits, mood prediction has its limitations. A person's mental condition is influenced by various factors such as genetics, environment, and life experiences, which may not be captured by the mood prediction algorithm. Especially, identifying external factors in the general case is still a hard task without a strong underlying theory. However, the use of ML might help in finding correlated variables - with multivariate analysis or principal component analysis;
    \item Need for Multimodal Mood Detection: Mood prediction and analysis can largely be improved by assessing information from different modalities (like physical appearance, questionnaires, and physiological data). However, it still poses the questions of the behaviour of a patient in an assessment compared to their behaviour in their daily situation, and the problem of available reliable devices.
\end{enumerate}

\section{Conclusion and Future Work}
\label{sec-conclusion}
In conclusion, in this paper, we pave the way for the creation of a questionnaire-based ML software for mood recognition and mood enhancement recommendation. Overall, the development of software like Psychlysis is an essential step toward addressing the increasing prevalence of mental health issues worldwide. The application provides a personalized approach to improving emotional well-being, aiding individuals, and mental health organizations in reducing the negative impact of mental health issues on daily life.

This \textbf{work-in-progress} application can build a profile of a user upon the personality traits analyzed from the user's answers to OCEAN and HEXACO inventories. With this information, the application can generate a questionnaire to assess user mood, predict factors that may impact positively or negatively user state of mind, and provide recommendations. To assist users to break away from ingrained patterns and behaviours, Psychlysis uses mood-altering practices like meditation, which have been found to have a good influence on a person's well-being. Data gathering, pre-processing, feature selection, and model creation are all part of the application's development process. The outcomes of the model preliminary evaluation show how well the application predicts the user's mood from personality traits and offers tailored recommendations.

However, we agree that there is still a lot of work to be done to achieve the creation of a fully operational system that can assist therapists and help users in doing concrete actions to improve their state of mind. One potential area is to address the inconsistency in the model's answers by exploring different approaches to train the model with larger and more diverse data sets. It may help in identifying multiple possible factors and hidden variables which impact user mood. Especially, it's important to assess the impact of specific mental troubles (like Bipolar Disorders, Personality Disorders...) on these variables compared to environments generating stress. By improving the consistency of the model's predictions, it may be possible to enhance the overall user experience and increase the effectiveness of the personalized recommendations. For that, it is necessary to make a huge interdisciplinary effort, including development process psychologists, psychiatrists, and neuroscientists to have a better description of personality traits that may impact the state of mind, create datasets correlating personality traits, and possibly positive/negative thoughts, and assess the adequacy between the predictions done by the software and the diagnosis of professionals in real therapy settings.

To improve accuracy, we can imagine this kind of application working with social media apps like Facebook and Instagram to get data from the user's feed activity in addition to getting it from a survey because, as we all know, social media is connected to everything these days and can help to provide recommendations based on search history. To train the machine learning model to make appropriate suggestions, the data gathered from social media applications may be more accurate. Moreover, new Large Language Models (LLM) like GPT-4 \cite{OpenAI:2023} can greatly contribute to making sure that questionnaires are well understood, by allowing the development of a real-time Chatbot interface.

Managing privacy, security and ethics when gathering, storing and processing private and sensitive information is also an important challenge when designing this kind of tool, in order to ensure that user data is protected and used responsibly by the different actors. Classical security techniques like encrypted storage, granular access rules and secured communication channels are indeed necessary to comply with regulations. Possible biases from datasets must be assessed carefully and kept in mind.

Finally, while the purpose of this preliminary work is the assessment of the potential of such an ML-based application for mood analysis using questionnaires, no validation protocol has been established to validate the results at a large scale. Therefore, strong and well-defined scientific validation protocol taking into account the diversity of people and situations must be defined in future work.
\bibliographystyle{IEEEtran}
\bibliography{IEEEfull,psychlysis}
\end{document}